\documentclass[twocolumn,preprintnumbers,amsmath,amssymb,showpacs]{revtex4}
\usepackage{graphicx}
\usepackage{dcolumn}
\usepackage{bm}
\usepackage{amssymb}
\usepackage{amsmath}
\usepackage{epstopdf}
\usepackage{color}
\usepackage{ulem}

\usepackage[colorlinks]{hyperref}

\begin{document}

\title{Dynamical structure factor of one-dimensional Bose gases: experimental signatures of beyond-Luttinger liquid physics}
\author{N. Fabbri*$^1$, M. Panfil$^{2,3}$, D. Cl{\'e}ment$^4$, L. Fallani$^{1,5}$,  M. Inguscio$^{1,5,6}$, C. Fort$^{1}$ and J.-S. Caux$^3$}
\affiliation{$^1$LENS European Laboratory for Non linear Spectroscopy, and Dipartimento di Fisica e Astronomia, Universit\`a di Firenze; INO-CNR Istituto Nazionale di Ottica del CNR, I-50019 Sesto Fiorentino, Italy \\
$^2$ SISSA-International School for Advanced Studies and INFN, Sezione di Trieste, I-34136 Trieste, Italy \\
$^3$Institute for Theoretical Physics, University of Amsterdam, Science Park 904, 1098 XH Amsterdam, The Netherlands \\
$^4$Laboratoire Charles Fabry, Institut d'Optique, CNRS, Univ. Paris Sud 11, F-91127 Palaiseau Cedex, France \\
$^5$QSTAR Center for Quantum Science and Technology, I-50125 Arcetri, Italy \\
$^6$ INRIM Istituto Nazionale di Ricerca Metrologica, I-10135 Torino, Italy
}

\begin{abstract}
Interactions are known to have dramatic effects on bosonic gases in one dimension (1D). Not only does the ground state transform from a condensate-like state to an effective Fermi sea, but new fundamental excitations, which do not have any higher-dimensional equivalents, are predicted to appear. In this work, we trace these elusive excitations via their effects on the dynamical structure factor of 1D strongly-interacting Bose gases at low temperature.  An array of 1D Bose gases is obtained by loading a $^{87}$Rb condensate in a 2D lattice potential. The dynamical structure factor of the system is probed by energy deposition through low-momentum Bragg excitations. The experimental signals are compared to recent theoretical predictions for the dynamical structure factor of the Lieb-Liniger model at $T > 0$. Our results demonstrate that the main contribution to the spectral widths stems from the dynamics of the interaction-induced excitations in the gas, which cannot be described by the Luttinger liquid theory.
\end{abstract}


\maketitle

\section{Introduction}
Most of our understanding of many-body quantum systems is anchored in the concept of (quasi)particles. Starting from free constituents obeying either fermionic or bosonic statistics, one builds a many-body ground state as either a Fermi sea or a Bose-Einstein condensate. In two or three dimensions, interactions then `adiabatically deform' the ground state into a Fermi liquid \cite{1957_Landau_JETP_3} with well-defined electron-like excitations, or a condensate state with Bogoliubov-like modes \cite{StringariBOOK}, respectively. In both cases, these well-defined excitations are conveniently described as quasiparticles and reveal themselves via sharp lines in dynamical response functions, indicative of free-particle-like coherently propagating modes.

Conversely, one-dimensional (1D) interacting systems are characterized by a breakdown of the basic Fermi liquid quasiparticle picture
\cite{GiamarchiBOOK}. The true quasiparticles of such systems are not adiabatically connected to free ones, and must be described using
a different language. This occurs because collective modes take over: applying perturbations does not create single Fermi liquid-like quasiparticles but rather an energy-continuum of excitations, so that the system response-functions develop features such as broad resonances and power-law asymptotes \cite{Dzyaloshinskii,1981HaldanePRL47,1963Lieb,1981Faddeev}.
In electronic systems and spin chains these characteristic continua of excitations have been recently observed \cite{Barak2010,Mourigal,Lake}, while their counterparts in interacting bosonic systems have not yet been unambiguously obtained.

In the last decades, cold atomic gases have offered a versatile experimental setting for studying different properties of 1D
systems (see \cite{Cazalilla2011} and references therein). More recently the study has been extended, {\it e.g.}, to the experimental investigation of thermodynamic properties \cite{Vogler2013}, impurity dynamics \cite{palzer2009, catani2012} and multiple spin fermionic systems \cite{pagano2014}.

In this work, we probe the excitations of strongly interacting 1D bosonic quantum gases at low temperature
using Bragg spectroscopy \cite{stenger1999,rpmDavidson}, in which a laser grating of amplitude $V$ imprints a perturbation onto the gas at specified momentum $\hbar q$ and energy $\hbar \omega$.
This results in a measurable increase of energy which depends on (and thus gives access to) the dynamical structure factor (DSF) $S(q,\omega)$ of the gas.
We present a new analysis of the Bragg excitations spectra previously reported in \cite{FabbriPRA11} through the comparison with a recent theoretical analysis of finite-temperature effects in the Lieb-Liniger model describing repulsively interacting bosons \cite{Panfil2013}. The excellent agreement with the solution of the model demonstrates that both quasiparticle and quasihole modes  with nonlinear dispersion relations are contributing to the experimental correlation signals, in a regime that lies beyond the reach of Luttinger Liquid theory.

The article is organized as follows. In Sec. \ref{Sec:Theo} we present the theoretical model used to describe the response of a single 1D gas to the Bragg excitation. In Sec. \ref{Sec:Exp} we describe the experimental procedure used to obtain the Bragg spectra. In Sec. \ref{Sec:Comp} we compare the theoretical and experimental findings, extending the theoretical model to account for the inhomogeneous distribution of tubes in the array. In Sec. \ref{Sec:Disc} we discuss the origin of the measured spectra, highlighting the role of interactions. Finally, In Sec. \ref{Sec:Concl} we summarize the results and present our conclusions.

\section{Theoretical description of a single interacting 1D gas at finite temperature.}
\label{Sec:Theo}

\begin{figure}[ht!]
\begin{center}
\includegraphics[width= \columnwidth]{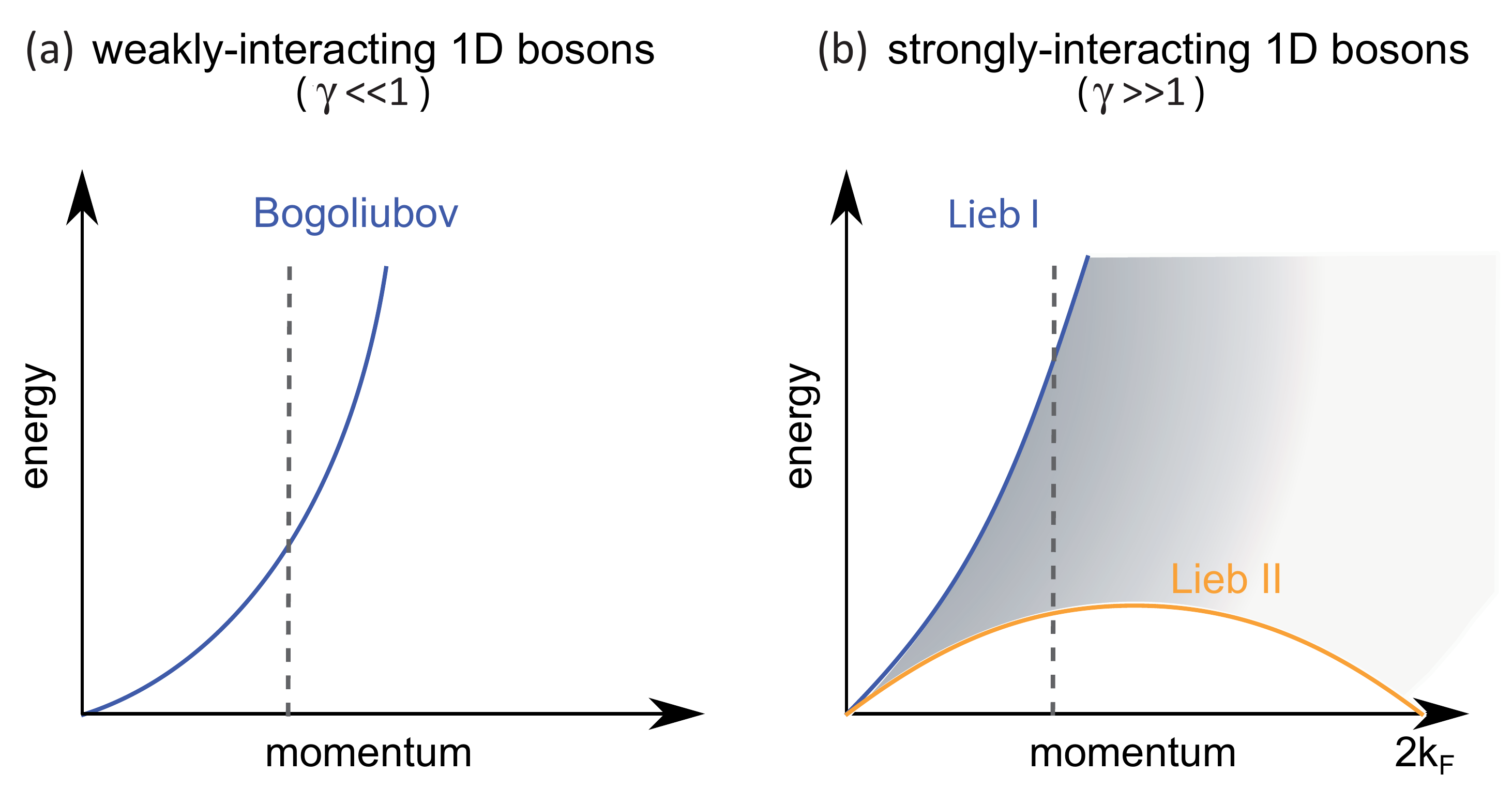}
\end{center}
\caption{Illustration of the effects of interactions on the excitation spectrum of a 1D gas. (a) Sketch of the excitation spectrum for a weakly-interacting Bose gas. (b) Sketch of the particle-hole spectrum for 1D strongly correlated bosons. The shaded area represents the continuum of excitations bounded between quasiparticle-like Lieb I mode and quasihole-like Lieb II mode. Gray dashed lines indicate cross-sections of the spectra at fixed momentum.
}
\label{Fig_sketch}
\end{figure}

A 1D Bose gas can be modeled by the  Lieb-Liniger Hamiltonian \cite{1963Lieb}
\begin{equation}
H = - \frac{\hbar^2}{2m} \sum_{j=1}^N \frac{\partial^2}{\partial^2_{x_{i}}} + g_{1D} \sum_{j>k=1}^N \delta(x_{j}-x_{k})
\label{Eq:LiebHamiltonian}
\end{equation}
where $m$ is the atomic mass and $g_{1D}$ is the strength of two-body contact repulsive interactions \cite{Olshanii1998}.
Two dimensionless parameters characterize the equilibrium state: the interaction strength $\gamma = m g_{1D} / (\hbar^2 \rho)$, $\rho$ being the atomic density, and the reduced temperature $\tau = 2m k_B T / (\hbar \rho)^2$ \cite{Kheruntsyan05}. Depending on the value of these parameters the equilibrium state of the 1D Bose gas resembles an ideal gas, a quasi-condensate or a Tonks-Girardeau gas \cite{PetrovPRL00,Sykes2008}.
While at $\gamma\ll 1$ the excitation spectrum is well described by Bogoliubov theory, at large enough $\gamma$ interactions create an effective ground-state Fermi surface with a Fermi wavevector $k_F = \pi \rho$, and the excitation spectrum is broadened into a continuum between Bogoliubov-like quasiparticles (Lieb I) and Fermi-like quasiholes (Lieb II) modes \cite{1963Lieb} as represented in Fig.~\ref{Fig_sketch}. At fixed momentum (gray dashed line in Fig. \ref{Fig_sketch} (a) and (b)), the energy spectrum in the two limiting cases will result in a single resonance for $\gamma\ll1$ and a broad asymmetric curve for $\gamma\gg1$ which tends to a uniform distribution bound between Lieb I and II modes for $\gamma\rightarrow\infty$ \cite{Caux2006}.

In the low but finite temperature regime of the experiment, $\tau \simeq 1$ and $\gamma \simeq 1$, both contributions from interactions and finite temperature have to be accounted for to quantitatively compare the experimental signals to the theoretical predictions. Computing dynamical correlations of the Lieb-Liniger model at finite temperature is a difficult task \cite{KorepinBOOK}, and a quantitative theory, for the specific case of the dynamical structure factor, has become available only recently \cite{Panfil2013} . This theoretical description allows us to understand and interpret the experimental spectra of an array of 1D gases, which was previously analyzed \cite{FabbriPRA11} with an incomplete theory based on the incorrect assumption of a mean-field regime where temperature was the major broadening source.

At finite temperature $T$, the increase of energy in the gas due to the Bragg pulse of time duration $t_B$ is proportional to $\omega S(q,\omega$) through the relation \cite{brunello2001}
\begin{equation}
\Delta E (q,\omega) = \left(\frac{2 \pi}{\hbar }\right)\left(\frac{V}{2}\right)^2 t_B \,\omega (1-e^{-\beta \hbar \omega}) S(q,\omega)
\label{DE}
\end{equation}
where $\beta=1/(k_BT)$. Using the above-mentioned theory \cite{Panfil2013}, we accurately evaluate the DSF for the Hamiltonian in  Eq.~(\ref{Eq:LiebHamiltonian}), {\it i.e.} for the integrable Lieb-Liniger model \cite{1963Lieb}. The method is based on the Bethe Ansatz, and relies on explicitly summing intermediate state contributions in a spectral representation.

\paragraph*{ Calculating the DSF.}
The dynamical structure factor can be expressed as
\begin{equation}
S(q, \omega) = \int \!dx \int_{-\infty}^\infty \!\!dt ~e^{i \omega t - i q x} \langle \hat{\rho}(x,t) \hat{\rho} (0,0) \rangle,
\label{Eq:DSF1}
\end{equation}
where $\hat{\rho}(x,t)$ is the density operator. In a Lehmann spectral representation, this becomes
\begin{equation}
S(q,\omega)= \frac{1}{\cal Z} \sum_{\lambda, \mu} e^{-\beta E_\lambda} |\langle \mu | \hat{\rho}_{q} | \lambda \rangle|^2 \ \delta (\omega- [E_{\mu}-E_{\lambda}]/\hbar),
\label{Eq:DSF}
\end{equation}
where ${\cal Z}$ is the partition function and the summations extend over all eigenstates of the system, $\hat{\rho}_{q}$ is the Fourier transform of the density operator $\hat{\rho}(x)$ and $E_{\mu}$ is the energy of an eigenstate $|\mu\rangle$, known through Bethe Ansatz \cite{1963Lieb}. Together with a knowledge of matrix elements of the density operator \cite{Slavnov1990} this allows for a precise numerical evaluation of the dynamical structure factor of a finite 1D Bose gas \cite{Panfil2013}. One particular aspect worth emphasizing here is that the matrix elements of the density operator are only non-negligible between states differing from each other by a very small (one or two) number of particle-hole (Lieb I and II) quasiparticle excitations; this means that the DSF (unlike more complicated observables) is sensitive to the presence of individual quasiparticles.

\begin{figure}[b!]
\begin{center}
\includegraphics[width= 1.0\columnwidth]{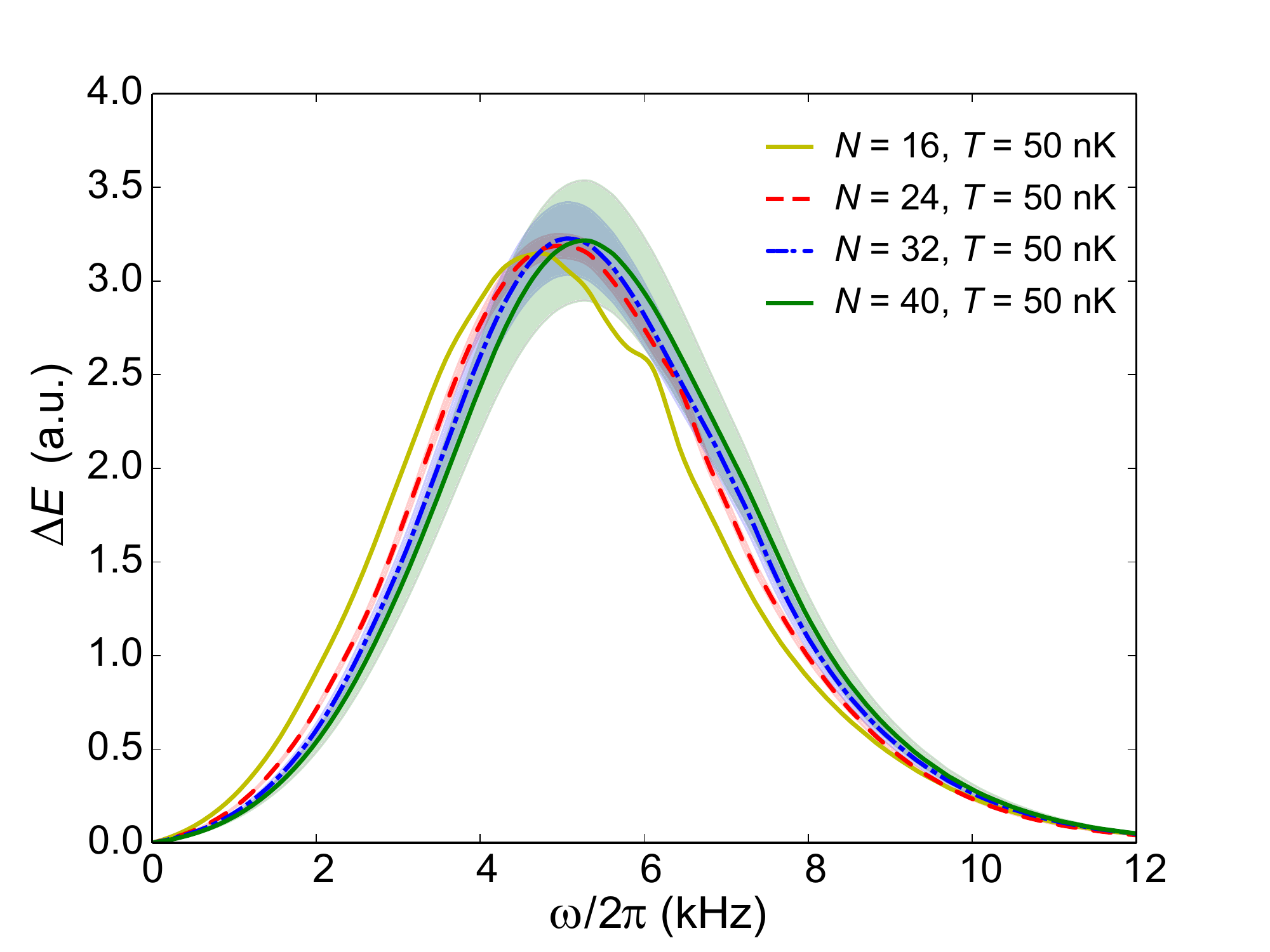}
\end{center}
\caption{Energy spectra calculated through thermodynamic Bethe ansatz for typical parameters of the experiment ($\rho=4.5$ $\mu$m$^{-1}$, and $ \gamma=1.2$) and different system sizes. The amplitude of the spectra is renormalized to have the same integral. The shaded areas represent the calculated uncertainty (see text). For system size $N\geq30$ the results are close enough to the thermodynamic one.}
\label{Fig_finitesize}
\end{figure}

In practice, the calculation method consists of the following steps. For a given interaction, density and temperature, a solution is found for the equilibrium state using the thermodynamic Bethe Ansatz \cite{1969_Yang_JMP_10}. A discrete system size $N$ is then chosen (as large as computationally practical) and a state is constructed which most closely matches the thermal one. The DSF is thereafter obtained by using the ABACUS algorithm \cite{2009_Caux_JMP_50} to sum the intermediate state contributions to $S(q,\omega)$ in order of decreasing importance, until satisfactory saturation of sum rule
\begin{equation}
\int_{-\infty}^\infty d\omega \ \omega S(q,\omega) =  N \frac{\hbar q^2}{2m},
\label{sumrule}
\end{equation}
is achieved.

\paragraph*{ Inhomogeneous trapped gas.}
The presence of a overall trapping potential along the axis of the 1D gas breaks the integrability of the Lieb-Liniger model. In order to compute the dynamical structure factor of a inhomogeneous trapped gas we first employ the local density approximation (LDA), where the response of the gas is assumed to be a sum of responses of small portions along the trap with different densities \cite{Kheruntsyan05}. We then verified that the response of the inhomogeneous gas is well approximated by the response of a uniform gas having a density equal to the mean density of the trapped one. The latter is used in the calculation.

\paragraph*{ Error estimation of the theoretical curves.}
Our calculation is affected by two types of inaccuracies. The first originates from lack of saturation of the f-sum rule, and it is represented by the shaded area in Fig. \ref{Fig_finitesize}. The second source of error comes from finite-size effects. Fig.~\ref{Fig_finitesize} presents the computation of the energy spectra at fixed density and for increasing system sizes, and shows that the response saturates for $N\geq 30$, well approximating the thermodynamic limit.

\begin{figure}[b!]
\begin{center}
\includegraphics[width= \columnwidth]{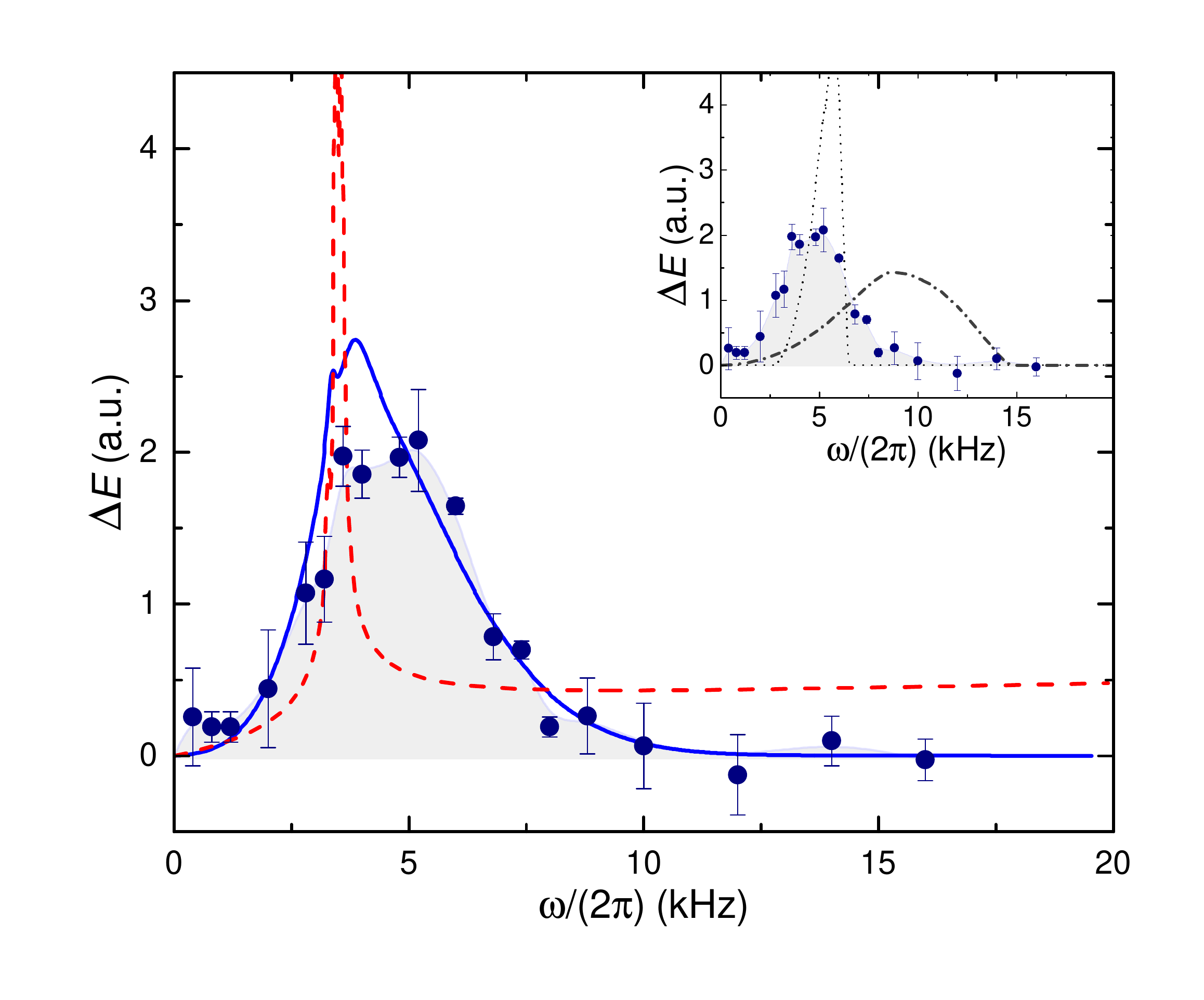}
\end{center}
\caption{Experimental data and theoretical spectra at {\it T}=0, illustrating the characteristics of different approaches (finite temperature effects will be discussed later). The graph shows the energy $\Delta E$ transferred to an array of 1D gases as a function of the Bragg excitation frequency $\omega$ for an excitation momentum $\hbar q \simeq 0.6 \hbar k_F$. The 1D gases are confined in a lattice of depth $s=35$. Blue dots: experimental data. The values are averages over up to 5 repeated measurements and the error bars are standard deviations. The shaded area below the experimental dots is a guide for the eyes.  Dashed line: energy transfer calculated according to the Luttinger liquid theory (see text). Solid thick line: Bethe Ansatz solution at interaction strength $\gamma\simeq1$ presented in this work. Inset: The same experimental data are shown along with the Bogoliubov theory for weakly interacting bosons ($\gamma \ll 1$, dotted line) and  Tonks-Girardeau limit for hard-core bosons ($\gamma=\infty$, dash-dotted line). The integral of each theoretical curve is normalized to that of the experimental signal. The conclusion is thus that a correct treatment of the interactions, as performed within Bethe Ansatz, already accounts for most of the observed interaction width.}
\label{Fig_limits}
\end{figure}

\section{Experimental Bragg spectra.}
\label{Sec:Exp}
We realize an array of independent 1D Bose gases, by loading a Bose-Einstein condensate of about $2\times10^5$ atoms of $^{87}$Rb in a two-dimensional optical lattice \cite{ClementPRL09}, produced by two laser standing-waves with wavelength $\lambda_L=830$nm aligned along the orthogonal directions $x$ and $z$.
The large amplitude of the optical lattices ($V_L = s \,E_r$ with $s = 35-50$, and where $E_r= h^2/2m \lambda_L^2$) results in very large radial trapping frequencies $\Omega_{\perp} / (2\pi) \simeq 40-50$ kHz. This plays a crucial role in reaching the regime of small temperatures $\tau \simeq 1$ as $\tau \propto \Omega_{\perp}^{-2}$, and $\gamma\simeq1$, where interactions affect the response of the system. The axial trapping frequency ranges from 53 to 63 Hz, when the lattice depth is varied from $s=35$ to $s=50$.

To probe the dynamical structure factor $S(q,\omega)$, we perturb the system with a Bragg pulse. In practice, two laser beams detuned by 200 GHz from the $^{87}$Rb D2 line are shone onto the atoms for a time duration $t_B=3$ ms, producing a Bragg grating potential $V \cos (q y - \omega t)$ where $V/h \simeq 900$ Hz \cite{linearity}.
The Bragg pulse induces two-photon transitions in the system. The transferred energy $\hbar \omega$ is varied by tuning the frequency difference of the two Bragg beams, whereas the momentum transfer is set by the fixed angle between the beams to $\hbar q = \hbar  \times 7.3(2)~\mu$m$^{-1}$, which corresponds to $q \simeq 0.6 k_{F}$ \cite{averagekF}.

The production of excitations is detected by monitoring the total energy $\Delta E$ deposited in the process as a function of $\omega$  \cite{ClementNJP09}.
In the experiment, after the excitation we decrease the lattice depth to $s=5$, letting the system thermalize for a few ms. We then measure the RMS size $\sigma$ of the central peak of the atomic density distribution after time-of-flight, which reflects the in-trap momentum distribution \cite{FabbriPRA11}.

We have verified the increase of $\sigma^2$ due to Bragg excitation, compared to the squared width $\sigma_{0}^2$ measured in the case of the unperturbed system, to be proportional to the increase of total energy \cite{ClementNJP09,FabbriPRA11}, and thus we can write
\begin{equation}
\sigma^2(\omega)-\sigma_{0}^2= \alpha \Delta E^{exp}(\omega).
\end{equation}
where $\alpha$ is a proportionality constant.

In Fig.~\ref{Fig_limits}  we report a typical experimental energy spectrum $\Delta E(\omega)$ (blue dots) with the energy transfer calculated according to the Luttinger liquid theory \cite{soundvelocity}, and the solution from the Bethe Ansatz approach presented in this work ($\gamma \simeq 1, \tau=0$, solid thick line). The comparison of these two theories with the experimental results enlightens that the Luttinger liquid model fails to describe the experimental results, whereas the Bethe ansatz solution of the Lieb-Liniger model successfully describes this intermediate regime of interactions where atomic dynamical correlations derive from both Lieb I and Lieb II modes. In the inset of Fig.~\ref{Fig_limits} we compare the same experimental data to two limiting cases where analytic predictions are available \cite{GolovachPRA09}: hard-core bosons in the Tonks-Girardeau regime ($\gamma = \infty, \tau=0$; dashed line), and weakly-interacting bosons \cite{Bogvelocity} ($\gamma \ll 1, \tau=0$, dotted line). These curves do not match the experimental findings, demonstrating that even if $\gamma\simeq1$ the experimental data show strong deviation from the weakly-interacting case. Note that although the $1/\gamma$ corrections to the DSF in the Tonks-Girardeau limit are known \cite{BrandCherny}, for the values of $\gamma \simeq 1$ they yield unphysical results. Therefore we consider only the limiting case of $\gamma=\infty$ for the purposes of comparison.

\section{Comparison between theory and experiment.}
\label{Sec:Comp}
Since in the experiment we have an inhomogeneous 2D array of 1D gases  where the atomic density $\rho_{k,l}$ varies across the array (each gas being labeled by indexes $k,l$), the total response is the sum of different lineshapes. To evaluate each contribution, the relevant  parameters of each gas are estimated as detailed in the following, and the contribution of each gas is weighted with its number of atoms $N_{k,l}$ since the DSF has to fulfil the f-sum rule and detailed balance.

\paragraph*{Atoms distribution in the array.}
As a first step, we estimate the distribution of atoms in the array when we load the 3D Bose-Einstein condensate in the 2D lattice potential. We use a rescaled interaction strength as proposed in \cite{KramerPRL02}.

The potential felt by the atoms consists in the sum of the 3D harmonic trapping potential and the optical potential from the 2D lattice. In the regime where the phase coherence is kept over the 3D cloud, ramping up the lattice mainly has two effects: {\it (i)} the local interaction parameter increases due to the tight confinement of the wells; {\it (ii)} the focusing of the lattice beams increases the overall trapping confinement along their transverse direction. These effects result in a change of the chemical potential $\mu_{{\rm3D}}$ and the trapping frequencies with the amplitude of the lattice $s$.

Assuming a Gaussian wave-function in the transverse directions, the first effect can be included by renormalising the interaction constant \cite{KramerPRL02},
\begin{equation}
\tilde{g}=g \frac{\pi}{2} \sqrt{s} \left( \frac{ {\rm Erf}[\pi s^{1/4}/\sqrt{2}]}{({\rm Erf}[\pi s^{1/4}/2])^2} \right)^2 \label{Eq:gRatio}
\end{equation}
where $g=4 \pi \hbar^2/a$, with $a$ the s-wave scattering length.

\begin{figure}[t!]
\begin{center}
\includegraphics[width= \columnwidth]{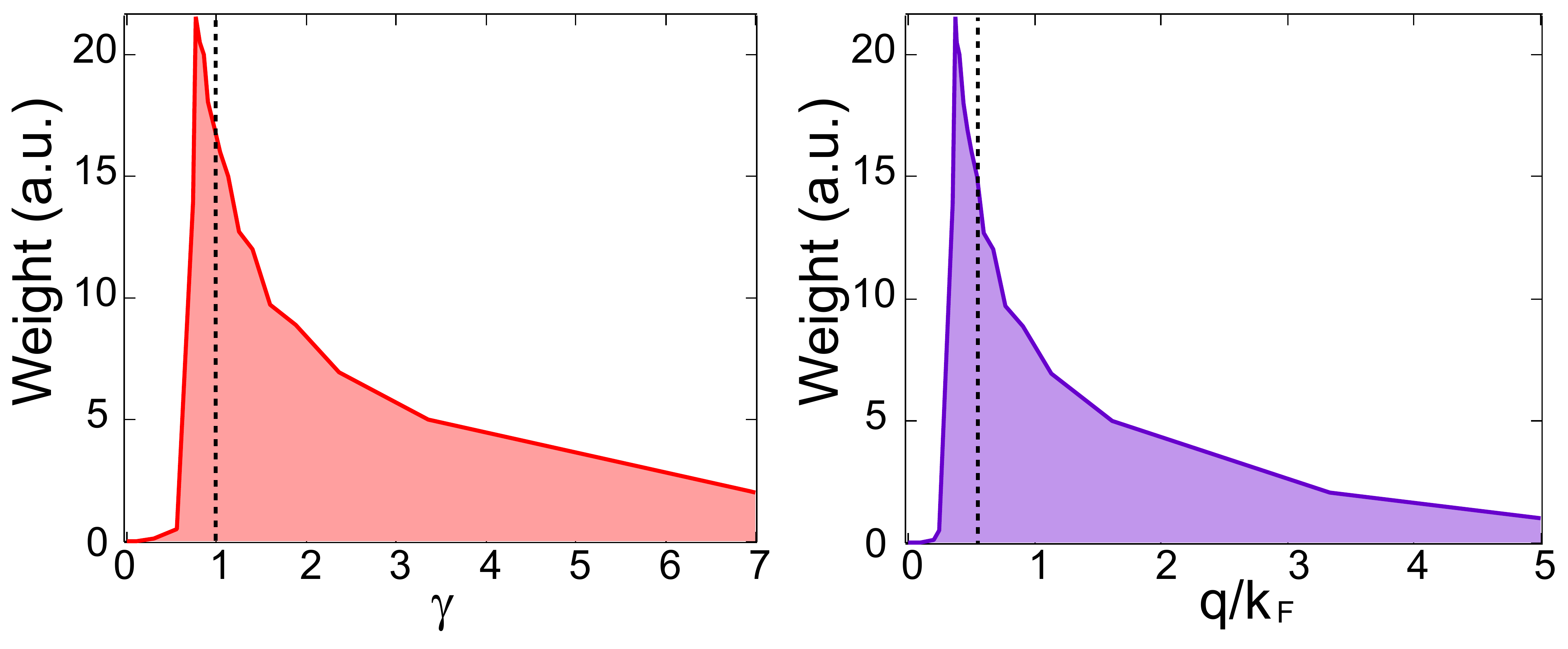}
\end{center}
\caption{Distributions of $\gamma_{k,l}$ (left) and $q/(k_{F})_{k,l}$ (right) in the 2D array of 1D gases as calculated from our model. The vertical dashed lines indicate the weighted mean values $\langle\gamma\rangle\simeq1$ and $\langle q/k_{F}\rangle\simeq0.6$.}
\label{Fig_gammadistrib}
\end{figure}

Concerning the change of the overall trapping frequencies, we make use of a harmonic approximation to evaluate the contribution from the optical lattice beams, an approximation that holds as long as the waist of the laser beam is larger than the size of the atomic cloud. The frequencies of the combined trap (magnetic and optical) are measured in the experiment through dipole oscillations. The results are in good agreement with the harmonic approximation.

\begin{figure*}[t!]
\begin{center}
\includegraphics[width= 1.6\columnwidth]{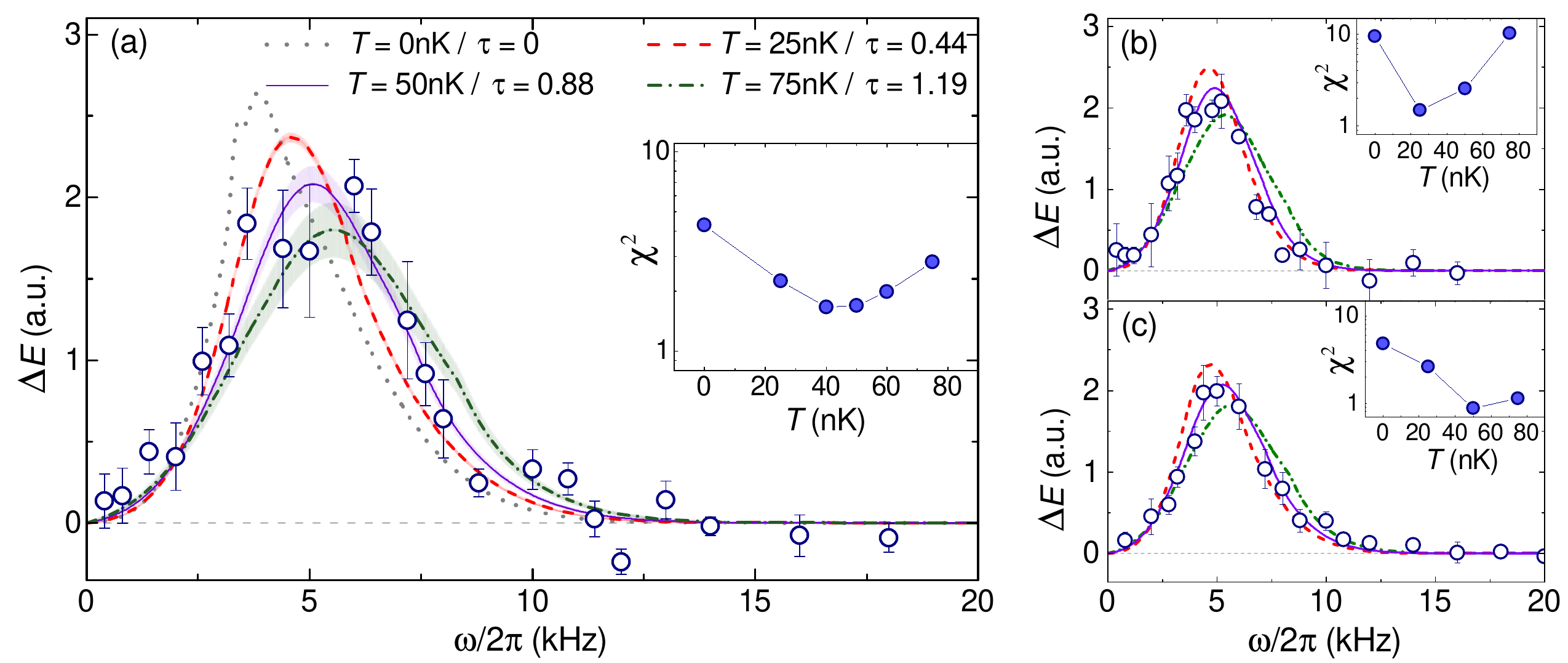}
\end{center}
\caption{Effect of finite temperature on the excitation spectra. The empty dots show the measured increase of energy of 1D gases confined in a 2D optical lattice of amplitude $s= 45$ (a), $s=35$ (b), and $s=50$ (c), after a Bragg excitation. The error bars are standard deviations over up to five repeated measurements per frequency. The lines are the calculated response of the array of 1D systems at different temperatures. For $s=35$, the temperatures $T=25$ nK, 50 nK, and 75 nK correspond to $\tau=0.57$, 1.15, and 1.72 respectively. For $s=50$, $\tau=0.39$, 0.79, and 1.18. The shaded area around the theoretical curves represents their uncertainty, mainly due to the normalization of the experimental data. The insets show the mean squared residuals between data and theoretical curves at different test temperatures $T$ ($\chi^2=1/N \sum_i (\Delta E_i^{e}-\Delta E_i^{t})^2$, where $\Delta E_i^{e}$ are the experimental data, $\Delta E_i^{t}$ the corresponding theoretical values, and $N$ the number of data-points).}
\label{Fig_exp}
\end{figure*}

Given the trapping frequencies, the chemical potential defines the overall Thomas-Fermi radius in the transverse directions $x$ and $z$, $R_{x,z}=\sqrt{2\mu_{{\rm 3D}}(s)/m \omega_{x,z}^2(s)}$.

The site indices $(k,l)$ range from $(0,0)$, which corresponds to the center of the trap, to $(k^{\rm M}_{x},k^{{\rm M}}_{z})$, defined as
\begin{eqnarray}
k^{{\rm M}}_{x,z} =  2 R_{x,z} / \lambda_L.
\end{eqnarray}

Assuming $k^{{\rm M}}_x=k^{{\rm M}}_z\equiv k_{{\rm M}}$ for simplicity of notation, the atom distribution can be written as
\begin{equation}
N_{k,l}=N_{0,0} \left(1- \frac{k^2+l^2}{k_{{\rm M}}^2} \right)^{3/2} \label{Eq:AtomDistrib}
\end{equation}
with $N_{0,0}=5 N/(2 \pi k_{{\rm M}}^2)$ the atom number in the central site. Similar atom distributions have been used to analyze the results of several experiments \cite{ParedesNature04,HallerScience09}.

Once the atom number distribution $N_{k,l}$ over the array of 1D gases has been calculated, we obtain the distributions of the interaction parameter $\gamma_{k,l}$ and the Fermi wavevector $(k_{F})_{k,l}=\pi \rho_{k,l}$.
In Fig.~\ref{Fig_gammadistrib} we plot the distribution of $\gamma_{k,l}$ and $q/(k_{F})_{k,l}$ for the experimental case.
The weighted mean values are typically $\langle \gamma \rangle \simeq 1$ and $\langle k_{F} \rangle \simeq q/0.6$, where the weight associated with a single gas $(k,l)$ is assumed to be equal to its relative atom number $N_{k,l}/N$ as justified by the f-sum rule in Eq. \ref{sumrule}.
From those distributions we calculate the energy spectra of each gas composing the 2D array.

\begin{figure}[t!]
\begin{center}
\includegraphics[width= \columnwidth]{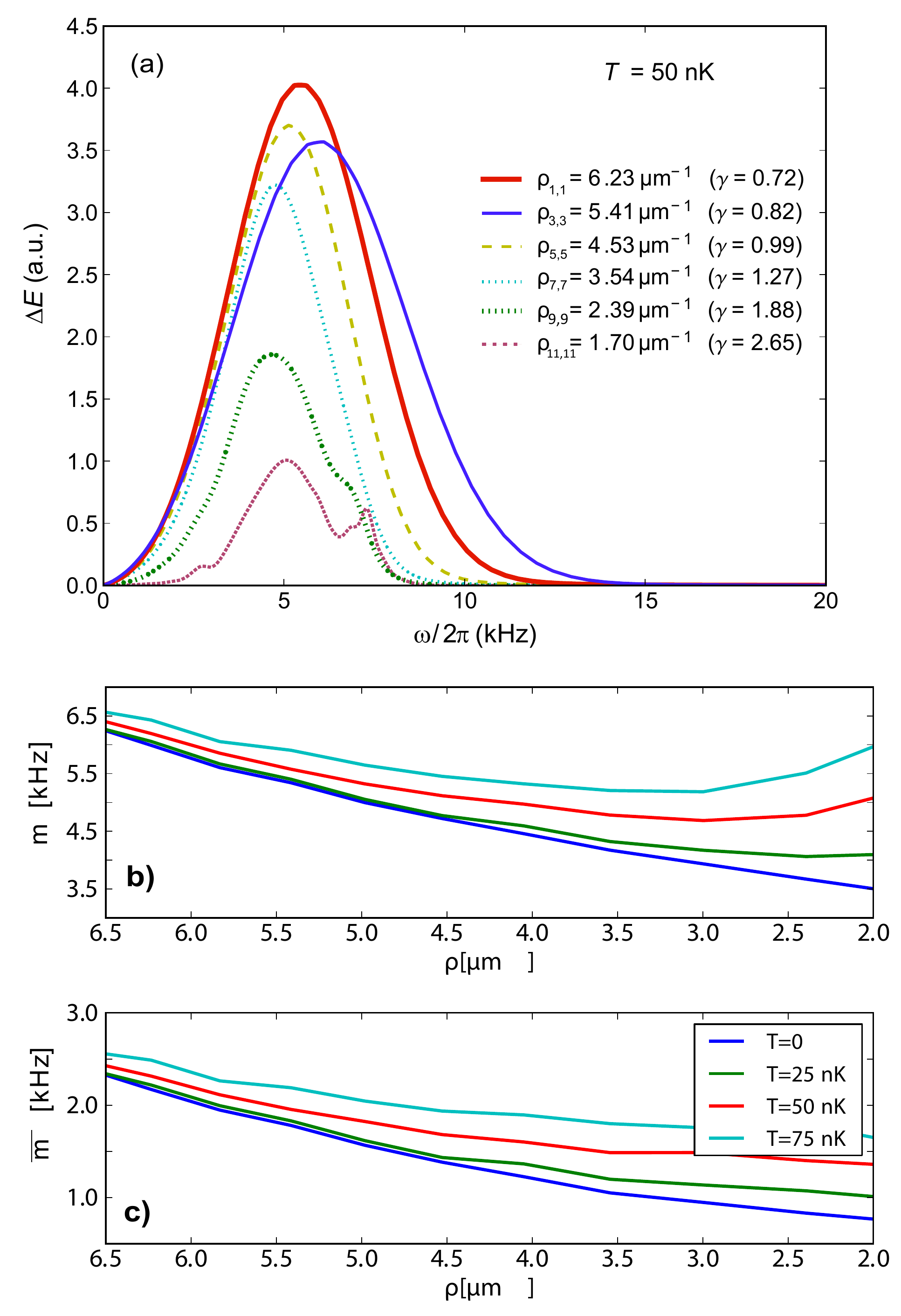}
\includegraphics[width= \columnwidth]{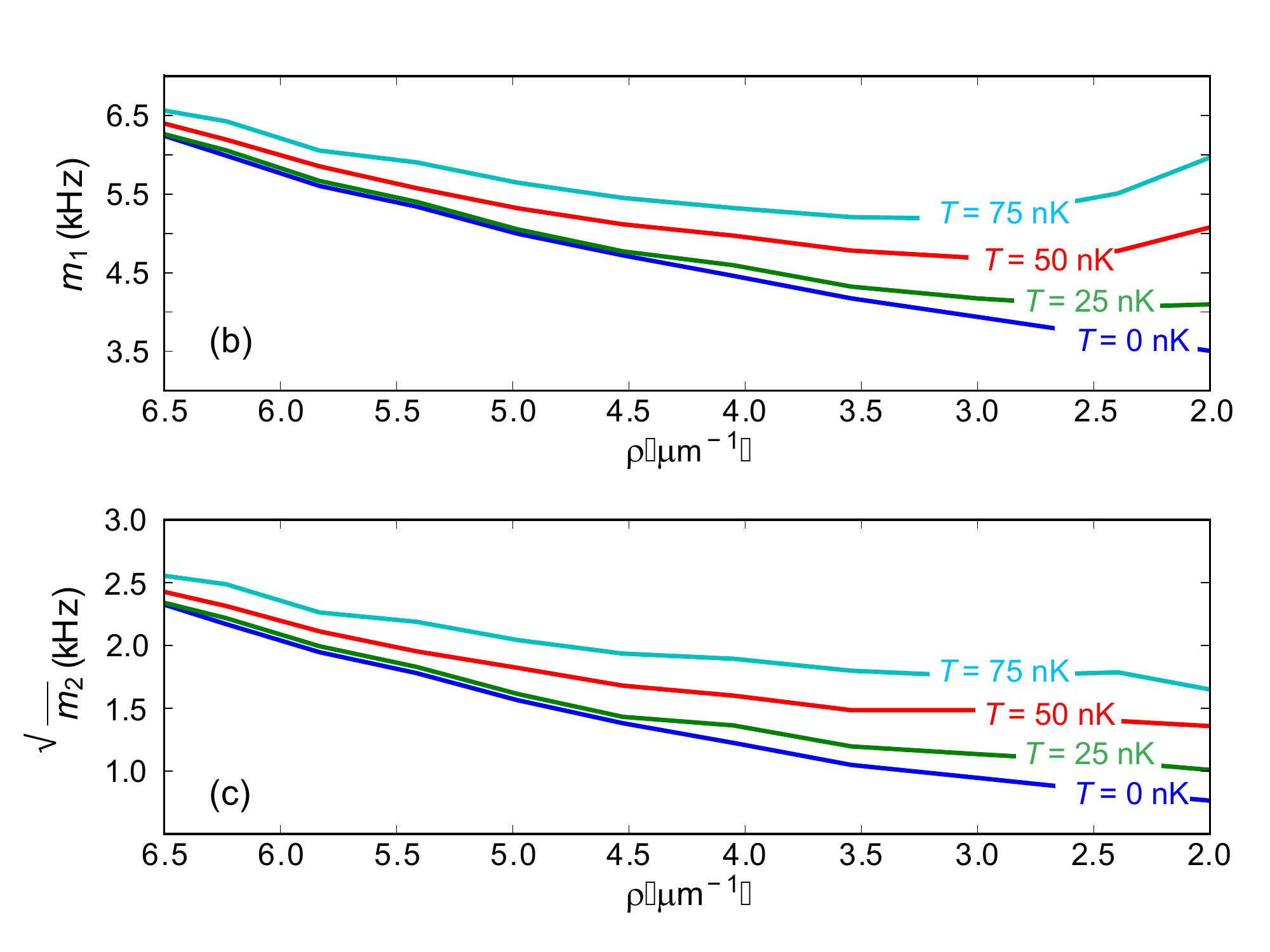}
\end{center}
\caption{Calculated energy spectra $\Delta E(q,\omega)$ for single 1D gases that compose the array. Atom distribution and parameters are those of the experimental spectrum of Fig.~\ref{Fig_exp} (a). (a) Calculated response of the single 1D gases of variable density $\rho_{k,l}$ composing the 2D array.
At low densities, finite-size effects are more pronounced than at high density, leading to artificial structures of the spectrum, noticeable at $\rho=1.70$ $ \mu$m$^{-1}$. (b)-(c) First two moments of $\Delta E(q,\omega)$ for a single 1D gas as a function of density $\rho$ and for different temperatures $T$.
}\label{Fig6}
\end{figure}

\paragraph*{Calculating the total response of the array of 1D gases.} Assuming the gases to be independent from each other during the Bragg excitation (the tunneling time between neighboring sites is about $\sim 200$ms at $s=35$, much larger than the duration of the Bragg excitation $t_{B}=3$ ms), we write the total response as the weighted sum of single gas responses.
Since averaging over the $\sim 2000$ different contributions that we estimate to have in the experiment would have resulted in impractically long computational times, we have evaluated the weighted sum on a representative set of 12 gases, whose parameters ($\rho$, $t$, $k_{F}$, ...) have been determined by coarse-graining the actual distribution.

\section{Origin of the measured spectral width.}
\label{Sec:Disc}

Figure \ref{Fig_exp} reports the comparison of the  experimental response of an array of 1D gases at $s=30,45,50$ with the Bethe-ansatz-based ABACUS predictions, which take into account all the experimental conditions as explained above, and is valid in the regime of interactions $\gamma\simeq 1$. Note that the calculation assumes temperature to be constant over the whole array \cite{thermalization}, and the theoretical and experimental data are normalized to their integral. The spectral broadening that can be observed in Fig. \ref{Fig_exp} has in principle different possible sources, {\it i.e.} inhomogeneity, temperature and interactions \cite{broadening}.

To analyze the contribution of inhomogeneity, Fig.~\ref{Fig6} (a) shows the calculated response of the single 1D gases that compose the array, weighted with their atom number. The central 1D gases, with larger atom numbers, dominate the total response: not only do they carry the largest weights,  but their response is also broad enough to cover that of the others.
Therefore the total response of the gas is well captured by considering a single `average' gas with $\langle \gamma \rangle \simeq1$,
$\langle \rho \rangle\simeq 5 ~\mu$m$^{-1}$ and $q/\langle k_{F}\rangle\simeq0.6$.
To get more physical insight, in Fig.~\ref{Fig6} (b)-(c) we plot the first two moments $m_1$ (average frequency) and $m_2$ (variance) of the calculated $\Delta E(q,\omega)$
as a function of the  density $\rho$ for different temperatures.
The scaling of the spectral width $\sqrt m_2$ with density, and therefore with $\gamma$, may seem inconsistent with common assertions about 1D Bose systems. Indeed, increasing $\gamma$ (at a fixed $q/k_F$ and $\omega_F$) is expected to lead to a broader frequency support of $S(q,\omega)$ and consequently of $\Delta E(q,\omega)\propto\omega S(q,\omega)$ \cite{Caux2006}. However, varying the density $\rho$ (at fixed $q$), as in the experiment, simultaneously sets {\it both} $\gamma$,  {\it and} the momentum and energy scales $k_F$ and $\omega_{F}$. Therefore, understanding the modifications in $\Delta E$ induced by a change of the density is  more involved than it is in the dimensionless approach.
A reduction of $\rho$ results in a larger interaction parameter $\gamma$ ($\propto \rho^{-1}$) and in a decrease of both  Fermi momentum ($\propto \rho$) and Fermi energy ($\propto \rho^{2}$). The differences in these scaling laws are responsible for the behavior of $m_{2}$ shown in Fig.~\ref{Fig6} (c), corresponding to a shrinking of the spectrum as decreasing the density.
Moreover at relatively high densities, as in our experiment, ($\rho \sim 5 ~\mu$m$^{-1}$)
both $m_1$ and $m_2$ only weakly depend on $T$ (for $\tau \lesssim 2$).
Then, we can conclude that the response of a single 1D gas has an energy width mostly due to interaction effects, rather than  finite-temperature broadening, and this conclusion can be extended to the whole array as confirmed by the direct comparison between experiments and theory in Fig.~\ref{Fig_exp}.

The large width of the measured spectra is evidence for the existence of quasiparticle- and quasihole- excitations (Lieb I and Lieb II modes) in a strongly-correlated Bose gas.
In contrast to the conclusions previously drawn by some of the Authors (when, in the absence of a complete theory at finite temperature,
they attributed large spectral widths to finite-temperature effects \cite{FabbriPRA11}), the present work demonstrates that interactions play a crucial role in our experimental observations while finite-temperature effects contribute only marginally. Nevertheless the analysis of the deviation of the experimental data from  the theory allows us to extract a best-fitting temperature of the order of $T\simeq40$ nK (see insets of Fig.~\ref{Fig_exp}).

\section{Concluding remarks.}
\label{Sec:Concl}
Fundamental excitations of strongly-correlated systems in one dimension, in contrast to their higher-dimensional counterparts, naturally lead to interaction-widened continua instead of sharp well-defined coherent lines. In our work, a new theoretical approach, which includes all the realistic conditions -- temperature, inhomogeneity and interactions -- allows us to interpret the experimental spectra of low-temperature interacting  1D gases and highlight the crucial role of interactions. In the regime of parameters of our experiment, with $\gamma\simeq 1$ and $\tau\simeq 1$, we have shown that the experimentally observed width in the dynamical structure factor, obtained using Bragg spectroscopy, can be ascribed to the underlying interactions instead of finite temperatures. The physical response can be understood from collective excitations of the Lieb-Liniger model that are not captured by the Luttinger liquid approach.
Our study validates this setup as an adequate quantum simulator for the further study of dynamical properties of in- and out-of-equilibrium 1D strongly-correlated bosons.

We thank M. Modugno for critical reading of the manuscript. M.P. and J.-S.C. acknowledge support from the FOM and NWO foundations of the Netherlands. D.C. acknowledges financial support from CNRS INP (collaboration LENS-LCF-LKB). N.F., D.C., L.F., M.I., C.F. acknowledge financial support from ERC Advanced Grant DISQUA, and EU FP7 Integrated Project SIQS.


\begin{thebibliography}{99}
\bibitem{linearity} We checked that the Bragg perturbation is in the linear response regime, by verifying experimentally the quadratic dependence of the observed signal on the Bragg grating amplitude $V$, as in Eq. \ref{DE}, up to twice the power used to obtain the spectra.
\bibitem{averagekF} We estimate $k_F$ as the average value overall the array.
\bibitem{Bogvelocity} We note that the Bogoliubov mean-field response is shifted toward higher frequencies than  the Luttinger response. This stems from the fact that the mean-field sound-velocity $v_s=\sqrt{g_{1D} \rho/m}$ (see Ref. \cite{GolovachPRA09}) is substantially larger than the Luttinger sound-velocity \cite{soundvelocity}. Moreover, the two expressions of sound velocity exhibit a different dependence on density, giving a different broadening due to the inhomogeneity of the array.
\bibitem{thermalization} Thermalization issue in 1D systems is by itself still under debate. In the absence of any concluding theory, and since we do not have experimental access to this parameter, we assume the whole array to be thermalized to the same temperature.
\bibitem{broadening} Other sources of broadening -- such as, that related to the finite frequency resolution of the Bragg pulse and the in-trap dynamics -- are negligible. The Bragg pulse duration ($t_B =$ 3 ms) corresponds to a frequency resolution of about 150 Hz, ten times smaller than the observed spectral width. This pulse duration is smaller than a quarter of the axial trapping period allowing us to neglect the effects from in-trap dynamics.
\end{thebibliography}
\end{document}